\documentclass[a4paper, 12pt]{article}
\usepackage{amssymb,amsmath}
\begin{document}

\thispagestyle{empty}

\title{New Learning and Testing Problems\\for Read-Once Functions}
\author{Andrey A. Voronenko\\Lomonosov Moscow State University\\Faculty of Computational Mathematics and Cybernetics}
\date{}

\maketitle

\begin{abstract}
In the paper, we consider several new types of queries for classical and new problems of learning and
testing read-once functions. In several cases, the border between polynomial and exponential
complexities is obtained.
\end{abstract}

\newpage

Classical function learning problem presupposes  consequent or parallel queries of function's values at points \cite{CHE218,HAN}. In most setups next queries can depend on the values of the function obtained with previous queries (a conditional test problem). The first classical problem of this type is monotone Boolean function learning, solved by G. Hansel \cite{HAN}. 
   
Consider a basis $B$. A function $f$ is said to be {\it read-once in the  basis $B$} iff there exists a  read-once formula in $B$ (a formula in which any variable occurs not more than once) expressing 
function $f$. In this paper we consider three bases of read-once functions:
$$
\{\land,\vee\},\ B_0=\{\land,\vee\,\neg\},\ B_2=\{\land,\vee,\oplus,\neg\}.
$$   
Read-once functions in the basis $\{\land,\vee\}$ are called {\it monotone read-once\\ functions}.
Read-once functions in the basis $B_0$ are simlpy called {\it read-once functions}. Such functions can
be uniquely represented as {\it trees} which have alternating levels of internal nodes labeled with $\land$ and $\lor$ and leaves labeled with literals $x_i$ and $\overline x_i$ (see, for example, \cite{HK}).

The tasks of learning and testing read-once Boolean functions often turn out difficult even to set up
non-trivially. It is natural to consider {\it degenerate} a problem with 
its answer being the number of the queries equal to $2^n$, which is the number of all Boolean vectors.

Bases for {\it read-once} (repetition-free) functions are usually considered\\  {\it hereditary}, i.e.,
if a basis contains some function, it must contain all its subfunctions obtained by constant substitutions. A basis is called {\it monolinear} iff it contains only monotone or only linear functions. The functions $0,1,x_1^{\sigma_1}\vee\dots\vee x_n^{\sigma_n},
x_1^{\sigma_1}\land\dots\land x_n^{\sigma_n}$ are read-once in any non-monolinear basis $B$ for any $\sigma_1,\dots,\sigma_n$. Even a {\it conditional diagnostical test} on such a set contains all queries.
Moreover, checking tests for constant functions also contain all Boolean vectors.
That's why the most-often considered read-once basis is $\{\land,\vee\}$.
The paper~\cite{AHK} suggests a solution to the conditional diagnostical test problem in this basis
using $O(n^2)$ queries. Other results can be found in  \cite{GUR203,GUR521,GOLDJ,ISR2005}.
Note that Russian papers almost always use the symbol $\&$ rather than $\land$ for denoting conjunction.

In  2002 the author  in \cite{MVK11} set up the problem of {\it testing with  respect to read-once alternatives} (denoted as  $Alt$ below). The problem is as follows. One has to check whether a given function is identical to a known function $f(x_1,\dots,x_n)$, which is  read-once in a basis  $B$ and depends essentially  on all its $n$ arguments. In addition, it is known a priori that a given function is  read-once in the basis  $B$
and depends only on the arguments $x_1,\dots,x_n$.
It was proved that for a basis $B_2=\{\land,\vee,\oplus,\neg\}$ the corresponding Shannon function equals
$1+n+\binom{n}{2}$ \cite{MVK11,RYAB}. Generalizations of this result to other bases and complexity characteristics can be found in  \cite{PM15,PM23,KAZ,PM33}.

In this paper we offer  new testing operations (queries).
On each step of an algorithm, a query addresses not a point, but a subcube of an arbitrary dimension.
If  the subcube dimension equals zero, we require that the result of the query  be equal to the value of the function at the corresponding point. The types of the queries are: 
\begin{trivlist}
\item $\oplus$ ---  the parity of the values of the function on the subcube;
\item $\neq$ --- $0$ if the function is constant on the subcube and $1$  otherwise;
\item $\sum$ ---  an arithmetic sum (the number of ones) of the Boolean values on the subcube.
\end{trivlist}

Denote a standard one-point query by $f(x)$ and a query on the subcube
$x_{i_1}={\sigma_1},\dots,x_{i_k}={\sigma_k}$ by
$F(-\dots-\sigma_1-\dots-\sigma_k-\dots-)$.
The first two types of subcube-oriented queries, like the usual one, give one bit of  output. The arithmetic sum query on an $m$-dimensional subcube can result in $m+1$ possible answers, thus giving $\left\lceil\log(m+1)\right\rceil$ bits of output.

In addition to testing with respect to read-once alternatives ($Alt$) and
conditional diagnostical test ($Diag$) problems mentioned above, we also consider a problem $Ess$ of constructing a conditional diagnostical test on the set of all read-once functions depending essentially on   $n$ variables. Degeneracy of this problem in the case of a non-monolinear basis is also obvious.
Denote by $L_T[P](M)$ the maximal complexity ({\it test length}, i.e., the
number of queries) of a problem $P$ with respect to a type of testing $T$ on a set of functions $M$.
Let $L_T[P,B](n)$ be equal to $L_T[P](M)$, where $M$ is  the set of all read-once functions in a basis $B$ which depend on $n$ variables ({\it Shannon function}).

{\bf Claim 1.} {\it For any query $T$ and problem $P$ the following inequality holds true:
$$
L_T[P,\{\land,\vee\}](n) \le L_T[P,B_0](n) \le L_T[P,B_2](n).
$$
}

{\bf Claim 2.} {\it For any problem $P$ and set of functions $M$ the following  inequalities hold true:
\begin{gather*}
L_{\sum}[P](M)\le L_{\oplus}[P](M)\le L_{f(x)}[P](M) \\
L_{\sum}[P](M)\le L_{\neq}[P](M)\le L_{f(x)}[P](M).
\end{gather*}
}

{\bf Claim 3.}  {\it For any query $T$ and any basis $B$ the following inequalities hold true:
\begin{gather*}
L_T[Alt,B](n)\le L_T[Diag,B](n) \\ L_T[Ess,B](n)\le L_T[Diag,B](n).
\end{gather*}
}

It may seem that for any set of functions $M$ and query $T$ the inequality
$L_T[Alt](M)\le  L_T[Ess](M)$ is true.
It turns out this is not true.

{\bf Example 1.} $L_{f(x)}[Ess,\{\land,\vee\}](2)=1$, but $L_{f(x)}[Alt,\{\land,\vee\}](2)=3$.

It is not difficult to choose a set of functions $M$ and a problem $P$ such that $L_{\oplus}[P](M)\le L_{\neq}[P](M)$.

{\bf Example 2.}
Consider the  set of seven functions, which depend on three variables:
the function $xy\vee xz\vee yz$ and six functions differing from it at  unique vectors with one or two ones. Any query of  type $\neq$ gives the same output for not less than  five functions. Therefore,
a diagnostical test cannot contain less than four queries. It is not difficult either to construct a conditional diagnostical test with queries  $\oplus$ asking at most three questions.

At the same time, there exists an example of an inverse relation for the same number of variables.

{\bf Example 3.}
Consider the following $32$ functions depending on variables $x_1,x_2,x_3$.
Eight of them are identically equal to zero on the subcube $x_3=0$ and have an odd number of ones on the subcube $x_3=1$.
Another eight  are identically equal to one on the subcube $x_3=0$ and have an odd number of ones on the subcube $x_3=1$.
Four of them are equal to $x_1\oplus x_2$ on the subcube $x_3=0$ and have a unique one on the subcube $x_3=1$.
Another four are equal to $x_1\sim x_2$ on the subcube $x_3=0$ and have a unique zero on the subcube $x_3=1$.
Finally, eight of them have an odd number of ones on the subcube $x_3=0$ and are identically equal to zero on the subcube  $x_3=1$.

It is not hard to construct a conditional diagnostic test of length $5$ using queries of  type $\neq$. Nevertheless, in  order to learn any set of $32$ Boolean functions with $5$ queries, one has to split the set in halves with each query. Among the queries of  type $\oplus$ only one-point queries $f(\alpha_1,\alpha_2,0)$ with  arbitrary constants $\alpha_1,\alpha_2$ satisfy this requirement. But there is no second query of  type $\oplus$ that can split any set of sixteen functions obtained on the first step into two sets of equal cardinality.

The relations $L_{f(x)}[Diag,B_0](n)=2^n$ and $L_{f(x)}[Ess,B_0](n)=2^n-1$ are trivial to prove.
It is proved in \cite{BUB} that $L_{f(x)}[Alt,B_0](n)=n+1$. In this paper we shall prove three results.

{\bf Theorem 1.} $$L_{\oplus}[Ess,B_0](n)=O(n^2). $$

{\bf Theorem 2.} $${1.5}^n\le L_{\oplus}[Diag,B_0](n)=O(n\cdot{1.5}^n). $$

{\bf Theorem 3.} $$L_{\sum}[Diag,B_0](n)=O(n^2). $$

The following fact is important for proving these theorems.

{\bf Lemma.} {\it Let $n\ge 1$. Suppose a  function $f(x_1,\dots,x_n)$ is read-once in the basis $B_0$.
Then it has an odd number of ones iff all its $n$ variables are essential.}

{\bf Proof.} An arbitrary function having an unessential argument evidently has an even number of ones.
If all the variables are essential, then by  definition of a read-once function the
following factorization holds true:
$$
f(x_1,\dots,x_n)=g(u_1,\dots,u_l)\circ h(v_1,\dots,v_m).
$$
If $\circ=\land$, then $|N_f|=|N_g|\cdot|N_h|$. If $\circ=\vee$, then
$|\overline{N_f}|=|\overline{N_g}|\cdot|\overline{N_h}|$.\hfill$\Box$

We  will also use a gradient cover lemma. The form given below and its proof can be found
 in the textbook \cite{LOZH}, p.~56.

{\bf Statement.} {\it Suppose that for some real $\gamma\  \left(0<\gamma\le 1\right)$ any column of a $0-1$ matrix $M$ of dimensions $p\times s$ contains not less than
$\gamma p$ ones. Then the cover of $M$ obtained with a gradient algorithm consists of no more than
 $\left\lceil\frac1{\gamma}\ln^{+}(\gamma s)\right\rceil+\frac1{\gamma}$ rows.}

Here $\ln^{+}x=\max\{\ln x,0\}$.

{\bf Claim 4.} {\it Suppose that a read-once function $f(y_1,\dots,y_t,u_1,\dots,u_p,v_1,\dots,v_q,w)$
depends essentially on all its arguments and the following equivalence holds:
$$
f(y_1,\dots,y_t,u_1,\dots,u_p,v_1,\dots,v_q,1)\equiv g(y_1,\dots,y_t,u_1\vee\dots\vee u_p\vee v_1\vee\dots\vee v_q).
$$
Also suppose that for some constants ${\bf c}$ the function $g(c_1,\dots,c_t,x)$ is identically equal to the variable $x$.
Then 
$$
f(y_1,\dots,y_t,u_1,\dots,u_p,v_1,\dots,v_q,w)=\psi(y_1,\dots,y_t,u_1\vee\dots\vee u_p\vee w\land(
v_1\vee\dots\vee v_q))
$$
iff for any $j$
$$
f(c_1,\dots,c_t,0,\dots,0,u_j=1,0,\dots,0,v_1=0,\dots,0,w=0)=1$$ and
$$f(c_1,\dots,c_t,0,\dots,u_p=0,0,\dots,0,v_j=1,0,\dots,0,w=0)=0.$$
Moreover, there is a constant $b$ such that for any $j$
$$
f(c_1,\dots,c_t,0,\dots,0,u_j=1,0,\dots,0,v_1=0,\dots,0,w=0)=b$$ and
$$f(c_1,\dots,c_t,0,\dots,u_p=0,0,\dots,0,v_j=1,0,\dots,0,w=0)=b$$
iff $f$ cannot be represented as
$$
f(y_1,\dots,y_t,u_1,\dots,u_p,v_1,\dots,v_q,w)=\psi(y_1,\dots,y_t,h(u_1,\dots,u_p,v_1,\dots,v_q,w))
$$
for any functions $\psi$ and $h$.
}

{\bf Proof of Theorem 1.} Consider the subcube $x_1=0$ query. By virtue of Lemma, if $F(0-\dots-)=0$, then $F(1-\dots-)=1$, and  if $F(0-\dots-)=1$, then $F(1-\dots-)=0$.
Then we proceed with learning the subcube with an odd number of ones and the variable $x_2$.
Note that 
$F(0-\dots-)=0$ or $F(1-\dots-)=1$ iff the variable  $x_1$ in {\it the tree} is connected with a conjunction or its negation $\overline{x}_1$ is connected with a disjunction. Similarly, $F(0-\dots-)=1$ or $F(1-\dots-)=0$ iff the variable  $x_1$ in {\it the tree} is connected with a disjunction or its negation $\overline{x}_1$ is connected with a conjunction. It is easy to learn a two-variable subfunction. Then we reconstruct all the subfunctions using Claim 4  for the variables $x_{n-2},\dots,x_1$ (in reverse order). The number of queries on each step is equal to the number of variables.\hfill$\Box$

{\bf Proof of Theorem  2.}

Lower bound. Consider the function $x_{i_1}^{\sigma_1}\land\dots\land x_{i_k}^{\sigma_k}$. Answer $1$ to a query will be obtained iff the requested subcube has a unique common point with the subcube
$x_{i_1}={\sigma_1},\dots,x_{i_k}={\sigma_k}$. Each subcube has a unique common point with exactly  $2^n$ subcubes. 
The number of all conjunctions of literals is $3^n$. Thus, if the number of queries is less than $1.5^n$, then  at least two different functions' answer sequences will be identically zero.

Upper bound. We will use Statement (gradient cover lemma).
Consider the matrix $M$ with rows and columns corresponding to subcubes of the cube $\{0,1\}^n$.
$m_{i,j}=1$ iff $i$-th and $j$-th subcubes have a unique common point.
So within our  conditions the parameters are:
$$
p=3^n,\ s=3^n,\ \gamma=\left(\frac23\right)^n.
$$
The power of the cover is not more than
$$
\left\lceil\left(\frac32\right)^n\ln\left(\left(\frac23\right)^n\cdot 3^n\right)\right\rceil+\left(\frac32\right)^n\sim
n\left(\frac32\right)^n\ln 2.
$$
The queries of the cover reveal all subcubes without unessential arguments, including the maximal one.
One needs no more queries to find this subcube. Now the problem is reduced to the problem
 $Ess$ for the basis $B_0$ and query type $\oplus$. It is sufficient to use Theorem 1 to estimate the needed complexity.\hfill$\Box$

{\bf Proof of Theorem  3.}
The equality $F(0-\dots-)=F(1-\dots-)$ means that the variable $x_1$ is unessential. In this case  the algorithm continues with any subcube of the last $n-1$ variables. In the opposite case we choose such a subcube that the number of ones on it has a one in the lowest possible binary digit.
Finally, we obtain a subcube with an odd number of ones and use Theorem~1.
 Note that in addition to the information  obtained in the process described in the proof of Theorem 1, here we on each step learn the monotonicity of a new variable.\hfill$\Box$

{\bf Mathematical aspects of physical models.}

We considered all queries to be of equal cost $1$. If $m$ is the dimension of a subcube beeing learned  and we obtain this information using usual queries, then the total cost of the subcube-oriented query equals  $2^n$.
However, some physical models can correspond to intermediate situations. Let the complexity of one query to an $m$-dimensional subcube be equal to $\phi(m)$ and  the number of such queries
be $\mu(m)$. 
We studied the value $\sum\limits_{m=0}^n\mu(m)$. Trivial one-point modeling costs
$\sum\limits_{m=0}^n2^m\mu(m)$. If we knew the value of $\phi(m)$, the needed complexity would be
represented as  $\sum\limits_{m=0}^n\phi(m)\cdot\mu(m)$.

We will now summarize the principal results known by now in the following table (one can use Claims 1--3 to obtain possible corollaries):

\bigskip

{\centering
\begin{tabular}{|c|c|c|c|}
\hline

&
$\{\land, \lor\}$
&
$\{\land, \lor, \neg\}$
&
$\{\land, \lor, \oplus, \neg\}$
\\\hline
$f(x)$
&
$L[Diag] = O(n^2)$
&
\parbox{3cm}{ $L[Diag] = 2^n$ \\ $L[Ess] = 2^n-1$ \\ $L[Alt] = n+1$ }
&
\parbox{3.5cm}{ $L[Diag] = 2^n$ \\ $L[Ess] = 2^n-1$ \\ $L[Alt] = \binom{n}{2}+n+1$ }
\\\hline
$\oplus$
&

&
\parbox{5.5cm}{ ${1.5}^n \leq L[Diag] \leq O(n \cdot {1.5}^n)$ \\ $L[Ess] = O(n^2)$ }
&

\\\hline
$\neq$
&

&

&

\\\hline
$\sum$
&

&
$L[Diag] = O(n^2)$
&

\\\hline
\end{tabular}
}

\bigskip

Obtaining accurate lower and upper bounds for the next three functions remains a significant unsolved problem:

\begin{enumerate}
\item $L_{\neq}[Ess,B_0](n)$.
\item $L_{\sum}[Diag,B_2](n)$.
\item $L_\oplus[Diag,B_0](n)$.
\end{enumerate}


\begin{thebibliography}{99}

\bibitem{AHK}  D.~ Angluin, L.~Hellerstein, M.~Karpinski.  Learning read-once formulas with queries.
\textit{Journal of ACM.} 1993. 40(1). 185--210.

\bibitem{BUB}  S.\,E.~Bubnov.  Funktsiya Shennona dliny proveryayushchikh testov funktsij, bespovtornykh v elementarnom bazise.
\textit{Sbornik statej molodykh uchenykh fakulteta VMiK MGU.} Vyp.~6. 47--57. [in Russian]

\bibitem{PM33}  S.\,E.~Bubnov, A.\,A.~Voronenko, D.\,V.~Chistikov.   Nekotorye otsenki  dlin testov dlya bespovtornykh funktsij v  bazise $\{\&,\vee\}$.
\textit{Prikladnaya matematika i informatika.} 2009. Vyp.~33. 90--100. [in Russian]

\bibitem{CHE218}  I.\,A.~Chegis, S.\,V.~Yablonskij.  Logicheskie sposoby kontrolya raboty elektricheskikh skhem.
\textit{Trudy matematicheskogo instituta imeni V.\,A.~Steklova.} Vol.~51. 1958. 270--360. [in Russian]

\bibitem{GOLDJ}  J.~Goldsmith, R.\,H.~Sloan, B.~Sz${\ddot o}$renyi, G.~Turan.  Theory revision with queries: horn, read-once, and parity formulas.
\textit{Artificial Intelligence.} 2004. V.~156. No.~2. 139--176.

\bibitem{ISR2005}  M.\,C.~Golumbic, A.~Mints, U.~Rotics.  Read-once functions revisited and the readability number of a Boolean function.
\textit{Electronic Notes in Discrete Mathematics.} 2005. Vol.~22. 357--361.

\bibitem{GUR203}  V.\,A.~Gurvich. Kriterij bespovtornosti funktsij algebry logiki.
\textit{Doklady AN SSSR.} 1991. Vol.~318. No.~3. 532--537. [in Russian]

\bibitem{GUR521}  V.\,A.~Gurvich. Repetition-free Boolean functions.
\textit{Uspekhi Matematicheskikh nauk.} 1977. Vol.~32. No.~1. 183--184. [in Russian]

\bibitem{HAN}  G.~Hansel.   Sur le nombre des fonctions bool$\acute e $ennes monotones de $ n $ variables.
\textit{C. R. Acad Sci. Paris}, 1966. Vol.~262. 1088--1090. [in French]

\bibitem{HK} L.~Hellerstein, M.~Karpinski. Learning Read-Once Formulas Using Membership Queries.
\textit{Proc. of the Second Annual Workshop on Computational Learning Theory}, Morgan Kaufmann Publishers, 1989. 146--161.

\bibitem{LOZH}  S.\,A.~Lozhkin. Lektsii po osnovam kibernetiki.
Moskva: 2004. Izd-vo fakulteta VMiK MGU. 256~s. [in Russian]

\bibitem{RYAB}  L.\,V.~Ryabets. Slozhnost proveryayushchikh testov dlya bespovtornykh bulevykh funktsij.
\textit{Irkutskij gosudarstvennyj pedagogicheskij universitet.} Ser. Diskretnaya matematika i informatika. 2007. Vyp.~18. 32~s. [in Russian]

\bibitem{PM15}  A.\,A.~Voronenko. Estimating the Length of a Diagnostic Test for some Nonrepeating Functions.
\textit{Computational Mathematics and Modeling.} 2004. Vol.~15. No.~4. 377--386.

\bibitem{MVK11}  A.\,A.~Voronenko. O proveryayushchikh testakh dlya bespovtornykh funktsij.
\textit{Matematicheskie voprosy kibernetiki.} Vyp.~11. Moskva: Fizmatlit, 2002. 163--176. [in Russian]

\bibitem{PM23}  A.\,A.~Voronenko. Recognizing the nonrepeating property in an arbitrary basis.
\textit{Computational Mathematics and Modeling.} 2007. Vol.~18. No.~1. 55--65. [in Russian]

\bibitem{KAZ}  A.\,A.~Voronenko, D.\,V.~Chistikov.   Individualnoe testirovanie bespovtornykh funktsij.
\textit{Uchenye zapiski Kazanskogo gosudarstvennogo universiteta.} Ser. Fiziko-matematicheskie nauki. 2009. Vol.~151. Book~2. 36--44. [in Russian]

\end{thebibliography}
\end{document}